\documentclass[aps,prl,
reprint,nofootinbib
]
{revtex4-2}
\usepackage{amssymb}
\usepackage[arrowdel]{physics}
\usepackage{amsmath}
\usepackage{amsthm}
\usepackage{tipa}
\usepackage{latexsym} 
\usepackage{graphicx}
\usepackage{slashed}
\usepackage{bbold}
\usepackage{cancel}
\usepackage{comment}
\usepackage{color}
\usepackage{graphicx,epsfig,color}
\usepackage{comment}
\usepackage{tikz}
\usepackage[compat=1.1.0]{tikz-feynman}

\newcommand{\vv}{``}

\begin{document}
\title{\Large{Physical Tuning} and Naturalness}
	
\author{Carlo Branchina}
\email{cbranchina@lpthe.jussieu.fr}
\affiliation{Laboratoire de Physique Th\'eorique et Hautes Energies (LPTHE),\\ UMR 7589,
	Sorbonne Universit\'e et CNRS, 4 place Jussieu, 75252 Paris Cedex 05, France.}
\author{Vincenzo Branchina}
	\email{branchina@ct.infn.it}
		\author{Filippo Contino}
	\email{filippo.contino@ct.infn.it}		
	\affiliation{Department of Physics, University of Catania, and INFN, \\
		Via Santa Sofia 64, I-95123 
			Catania, Italy	}

\begin{abstract}
We present a radically new proposal for the solution of the naturalness/hierarchy problem, where the fine-tuning of the Higgs mass  finds its physical explanation and
the well-known multiplicative renormalization of the usual perturbative approach emerges as an IR property of the non-perturbative  running of the mass.

\end{abstract}
	
\maketitle
	
The Standard Model (SM) of particle physics is an effective theory, i.e.\,a quantum field theory valid up to a certain scale ($M_P$, $M_{GUT}$, ...), above which it has to be replaced by its ultraviolet (UV) completion. This scale, that we generically indicate with $\Lambda$, is the {\it physical cut-off} \,of the theory: the SM effective Lagrangian 
${\mathcal L}_{SM}^{(\Lambda)}$,
allows to describe processes at momenta $p \lesssim \Lambda$.   

Due to unsuppressed quantum fluctuations, the square of the Higgs boson mass $m^2_H$ receives contributions proportional to  $\Lambda^2$. In this respect, we stress  that $m^2_H \sim \Lambda^2$ indicates a \vv quadratic sensitivity'' of $m^2_H$ to the ultimate scale 
of the theory, not a \vv quadratic divergence''. Moreover, this value of $m^2_H$ is nothing but the square of the running $m^2_H(\mu)$ at the scale $\mu=\Lambda$. If $\Lambda$ is too large, $m^2_H(\Lambda)$ is \vv unnaturally" large, and this poses a problem of \vv hierarchy''  with the Fermi scale $\mu_F$,
where $m_H(\mu_F)\sim 125$ GeV. Several attempts have been made towards the \vv solution'' of this naturalness/hierarchy (NH) problem. Here we focus on three of them, as they will help to introduce our proposal.  
  
1. A popular approach is based on the assumption that the UV completion of the SM could provide  the condition
\begin{equation}\label{natur}
m^2_H (\Lambda) \ll \Lambda^2\,
\end{equation}
at the scale $\Lambda$\,\cite{Giudice:2013yca,Holthausen:2013ota,Chankowski:2014fva}.
Sometimes (\ref{natur}) is viewed as a \vv quantum gravity miracle''\cite{Giudice:2013yca}, that could result from a conspiracy among the SM couplings at the scale $\Lambda$. This is for instance the case of the so called Veltman condition\footnote{If not directly to the SM itself, it can be applied to some of its extensions\,\cite{Chankowski:2014fva}.}\,\cite{Veltman:1980mj}.
In such a scenario: 
(i) the naturalness problem is solved from physics \vv outside'' the SM realm, as (\ref{natur}) is considered a left-over of its UV completion (or extensions of it); 
(ii) the hierarchy problem is solved \vv inside'' the SM, by considering the perturbative renormalization group (RG) equation for $m_H^2(\mu)$ ($\gamma \ll 1$ is the perturbative anomalous mass dimension)
\begin{align}\label{submass}
\mu\frac{d}{d\mu} m_H^2(\mu)&=\gamma\,\, m_H^2(\mu)\,.
\end{align}

From (\ref{natur}) and (\ref{submass}), in fact, $m^2_H(\mu_F)$ and  $m^2_H(\Lambda)$ turn out to be of the same order: there is no problem of hierarchy. 

2. A somehow complementary approach consists in considering again Eq.(\ref{submass}) for $m^2_H(\mu)$, but assuming this time that gravity could provide a non-perturbative value for $\gamma$ ($\sim 2$). In this case, the large hierarchy between the Fermi scale $\mu_F$ and the UV scale $\Lambda$ can be accommodated\,\cite{Bornholdt:1992up,Pawlowski:2018ixd,Wetterich:1981ir,Wetterich:1983bi,Wetterich:1990an,Wetterich:1991be,Wetterich:2011aa,Wetterich:2016uxm}: the NH problem would then disappear. 

3. Finally, some authors suggest that dimensional regularization (DR) could be endowed with special physical  properties that  make it the correct \vv physical'' way to calculate the radiative corrections in quantum field theory (QFT). Again, if no new heavy particles are coupled to the Higgs boson, the NH problem would seem to be absent from the  beginning\,\cite{Meissner:2006zh,Salvio:2014soa,Boyle:2011fq,Alexander-Nunneley:2010tyr,Carone:2013wla,Farzinnia:2013pga,Ghilencea:2015mza,Ghilencea:2016dsl,Guo:2014bha,Kawamura:2013kua,Foot:2007iy,Meissner:2007xv,Oda:2018zth,Ghilencea:2016ckm,Heikinheimo:2013fta,Mooij:2018hew,Shaposhnikov:2008xi,Bezrukov:2014ipa,Bezrukov:2007ep,Bars:2013yba,Farina:2013mla,Brivio:2017vri,Steele:2013fka,Wang:2015cda}.

As shown in a recent paper\,\cite{Branchina:2022jqc},\,however, none of these approaches provides a solution to the  problem. The reason is that any effective field theory (EFT), including the SM, is {\it necessarily} defined and interpreted in a Wilsonian framework. The meaning of this statement is twofold: (i) the parameters (masses and couplings) $g_i(\Lambda)$ in the effective Lagrangian ${\mathcal L}_{SM}^{(\Lambda)}$ result from integrating out the higher energy modes\, $k > \Lambda$ \,related to the UV completion of the SM; (ii) the same  parameters $g_i (\mu)$ at a lower scale $\mu < \Lambda$ result from integrating out the modes of the fields that appear in ${\mathcal L}_{SM}^{(\Lambda)}$ in the range $[\mu, \Lambda]$.  
 
It is shown in\,\cite{Branchina:2022jqc} that DR provides a specific implementation of the Wilsonian strategy, where the fine-tuning is {\it automatically} encoded in the calculations, although in a  {\it hidden} manner. As a consequence, DR cannot provide a solution to the problem. Moreover, it is shown that (\ref{submass}) is obtained when the \vv critical value'' $m^2_{cr}(\mu)$
is subtracted to $m^2(\mu)$. 
In other words,   $m_H^2(\mu)$ in (\ref{submass}) is not the Wilsonian mass  $m^2(\mu)$. It is rather: $m_H^2(\mu)\equiv m^2(\mu) - m^2_{cr}(\mu)$.   Equation \,(\ref{submass}) then  incorporates the fine-tuning, and cannot be invoked to solve the NH problem. 

In this {\it Letter} we make a radically different proposal, rooted in simple and (in our opinion) indisputable \vv facts'': (i) the SM is an EFT valid up to an ultimate UV scale $\Lambda$;  
(ii) the Wilsonian  integration of modes is the only {\it physically consistent} way of including the quantum fluctuations in an EFT.
	
To introduce our proposal, we begin by considering the Wilsonian RG equations for the scalar $\phi^4$ theory, whose (Euclidean) Lagrangian is
\begin{equation}
	\mathcal L=\frac{1}{2}\partial_\mu\phi\partial^\mu\phi+\frac 12 m^2_{_{\Lambda}}\phi^2 +\frac{\lambda_{_{\Lambda}}}{4!}\phi^4,
\end{equation}
where $m^2_{_{\Lambda}}\equiv m^2(\Lambda)$ and $\lambda_{_{\Lambda}}\equiv \lambda(\Lambda)$ are the mass and coupling constant at the (physical) scale $\Lambda$. 
By considering the corresponding Wilsonian action within the so called \vv Local Potential approximation",
$S_k[\phi]=\int d^4\,x \left(\frac{1}{2}\,\partial_\mu\phi\,\partial_\mu\phi+ U_k(\phi)\right)$, 
and truncating the potential to the first two terms, $U_k(\phi)=\frac 12 m^2_{_{k}}\phi^2 +\frac{1}{4!}\lambda_{_{k}}\phi^4$, the RG equations for $m^2_k$  and $\lambda_k$ are (see \cite{Bonanno:1999ik,Branchina:2022jqc})
\begin{align}
	\label{mk}
	k \frac{d m^2_k}{d k}&=-\frac{k^4}{16\pi^2}\frac{\lambda_k}{k^2+m^2_k}, \\
	\label{lambdak}
	k \frac{d \lambda_k}{d k}&=\frac{k^4}{16\pi^2}\frac{3 \lambda_k^2}{(k^2+m_k^2)^2} .
\end{align}

When the UV boundaries for  (\ref{mk})-(\ref{lambdak}) are such that  the condition $m^2_k \ll k^2$ is satisfied in the whole range of integration, this system is well approximated by
\begin{eqnarray}
	\label{runningeqm}
	k\frac{dm^2_k}{dk}&=&-\frac{\lambda_k}{16\pi^2}k^2+\frac{\lambda_k}{16\pi^2}m_k^2\\
	\label{runningeql}
	k\frac{d\lambda_k}{dk}&=&\frac{3\lambda_k^2 }{16\pi^2}\,.
\end{eqnarray}
Taking for instance \vv SM-like" IR boundaries,  $m(\mu_F)=125.7$ GeV and $\lambda(\mu_F)=0.1272$, and solving both systems numerically, we find that the solutions to (\ref{mk})-(\ref{lambdak}) and  (\ref{runningeqm})-(\ref{runningeql}) coincide with great accuracy.

The flow equations (\ref{runningeqm})-(\ref{runningeql}) can be solved analytically. The solution to  (\ref{runningeql}) (decoupled from (\ref{runningeqm})) 
is the well-known one-loop-improved running quartic coupling 
\begin{equation}\label{runninglambda}
	\lambda(\mu)=\frac{\lambda_{_{\Lambda}}}{1-\frac{3}{16\pi^2}\lambda_{_{\Lambda}}\log\left(\frac{\mu}{\Lambda}\right)}   .
\end{equation}
Then, inserting (\ref{runninglambda}) in (\ref{runningeqm}), this latter equation can also be solved 
analytically, and we find 
\begin{align}\label{runningmass2}
	&m^2(\mu)=\frac{1}{3\cdot2^{2/3} \left(3 \lambda_{_{\Lambda}}\log \left(\frac{\mu }{\Lambda }\right)-16 \pi ^2\right)}\Bigg(2^{2/3} \Lambda ^2 e^{\frac{32 \pi ^2}{3 \lambda_{_{\Lambda}} }} \times \nonumber \\
	&\left(16 \pi ^2-3 \lambda_{_{\Lambda}}  \log \left(\frac{\mu }{\Lambda }\right)\right) E_{\frac{2}{3}}\left(\frac{32 \pi ^2}{3 \lambda_{_{\Lambda}} }-2 \log \left(\frac{\mu }{\Lambda }\right)\right)\nonumber\\
	&+4 \lambda_{_{\Lambda}}  \sqrt[3]{-\frac{1}{\lambda_{_{\Lambda}} }} \left(\Lambda ^2 e^{\frac{32 \pi ^2}{3 \lambda_{_{\Lambda}}}} E_{\frac{2}{3}}\left(\frac{32 \pi ^2}{3 \lambda_{_{\Lambda}}}\right)+3 m^2_{_{\Lambda}}\right) \nonumber\\
	&\,\,\times \left(3 \pi  \log \left(\frac{\mu }{\Lambda }\right)-\frac{16 \pi ^3}{\lambda_{_{\Lambda}} }\right)^{2/3}\Bigg)
\end{align}
where $E_{\frac23}(x)$ is the generalized exponential integral function $E_{p}(x)$ with $p=\frac23$.

The {\it non-perturbative}
evolution equation that we have just  found is important for our analysis. First of all we note that, expanding  (\ref{runningmass2}) for $\lambda_{_\Lambda}\ll1$ (and $\mu^2 \ll \Lambda^2$), we obtain the well-known perturbative result (for notational simplicity,  from now on we replace $\lambda_{_\Lambda} \to \lambda$)
\begin{align}m^2_\mu=m^2_{_\Lambda} + \frac{
\lambda}{32\pi^2}\left(\Lambda^2- m^2_{_\Lambda} \log\left(\frac{\Lambda^2}
{\mu^2}\right)\right)\,.
\end{align} 
More important for our scopes, however, is to note that the flow equation (\ref{runningmass2}) has a very interesting {\it non-perturbative} approximation, that can be obtained  replacing $\lambda_k$ with $\lambda$ in the right hand side of (\ref{runningeqm}), 
\begin{equation}\label{riscrittura}
	m^2_\mu=\left(\frac{\mu}{\Lambda}\right)^{\frac{\lambda}{16\pi^2}}\left(m^2_{_{\Lambda}}+\frac{\lambda \Lambda^2}{32\pi^2-\lambda}\right)-\frac{\lambda\mu^2}{32\pi^2-\lambda}\,. 
\end{equation}
Using for instance the same boundary values considered  before (see below (\ref{runningeql})), we easily check that  (\ref{riscrittura}) provides a very good approximation to the flow governed by (\ref{mk}).

Eq.\,(\ref{riscrittura}) is a crucial result of the present work (below it will be extended to the SM) and contains several important lessons.  First it shows how the fine-tuning usually realized in perturbative QFT  operates in the Wilsonian framework: it simply fixes the boundary at the UV scale $\mu=\Lambda$ for  the running of the mass $m^2_\mu$. This is encoded in the parenthesis in the right hand side of  (\ref{riscrittura}), where 
$m^2_{_\Lambda}$ and $\frac{\lambda \Lambda^2}{32\pi^2-\lambda}$ need to be {\it enormously fine-tuned} if at the IR scale $\mu_{low}$ we want  $m_{{\mu_{low}}}\sim\mathcal O (100)$ GeV.

There is another important lesson in (\ref{riscrittura}). By simple inspection, we see that the combination
\begin{equation}\label{mtilde}
m_{\mu,r}^2\equiv m^2_\mu+\frac{\lambda \,\mu^2}{32 \pi^2-\lambda}
\end{equation}
obeys the RG
equation 
\begin{equation}\label{subtraction}
	\mu \frac{d}{d \mu}  m_{\mu,r}^2= \gamma \,  m_{\mu,r}^2
\end{equation}
where $\gamma=\frac{\lambda}{16\pi^2}$
is the mass anomalous dimension for the $\phi^4$ theory at one-loop order.
Eq.\,(\ref{subtraction}) coincides with the well-known one-loop improved flow equation for the renormalized running mass. Therefore, $m_r^2(\mu)$ defined in (\ref{mtilde}) {\it is} the renormalized running mass. 
At the same time we note that 
\begin{equation}\label{scalarMC}
	m^2_{\mu, cr} \equiv - \frac{\lambda\,\mu^2}{32 \pi^2-\lambda}
\end{equation}
is the  \vv critical mass" defined at each value of the running scale $\mu$, and that the  subtraction in (\ref{subtraction}) drives the RG flow close to the critical surface of the Gaussian fixed point. The simple integration of (\ref{subtraction}) gives ($\mu_0\, > \mu$) 
\begin{equation}\label{solpert_RG}
	m^2_{\mu, r}=\left(\frac{\mu}{\mu_0}
	\right)^{\frac{\lambda}{16\pi^2}}
	m^2_{\mu_0, r}\,.
\end{equation}

For the purposes of our analysis, it is important to note that we derived  equation\,(\ref{subtraction}) in the Wilsonian framework, namely from the RG flow (\ref{riscrittura}), whereas usually it is derived in the context of \vv technical schemes", as dimensional, heat kernel, or zeta function regularization.
In this respect, we stress that, when the quantum fluctuations are calculated in the framework of a \vv technical scheme", we only have access to (\ref{subtraction}) (and then to its solution  (\ref{solpert_RG})), but we are blind to the fact that the renormalized running mass is obtained only after operating   at each scale $\mu$ the subtraction in (\ref{mtilde}). 
When, on the contrary, the quantum fluctuations are calculated within the Wilsonian \vv physical scheme", we clearly see how the  renormalized mass emerges. 

\begin{figure*}[t]
	\centering
	\includegraphics[scale=0.41]{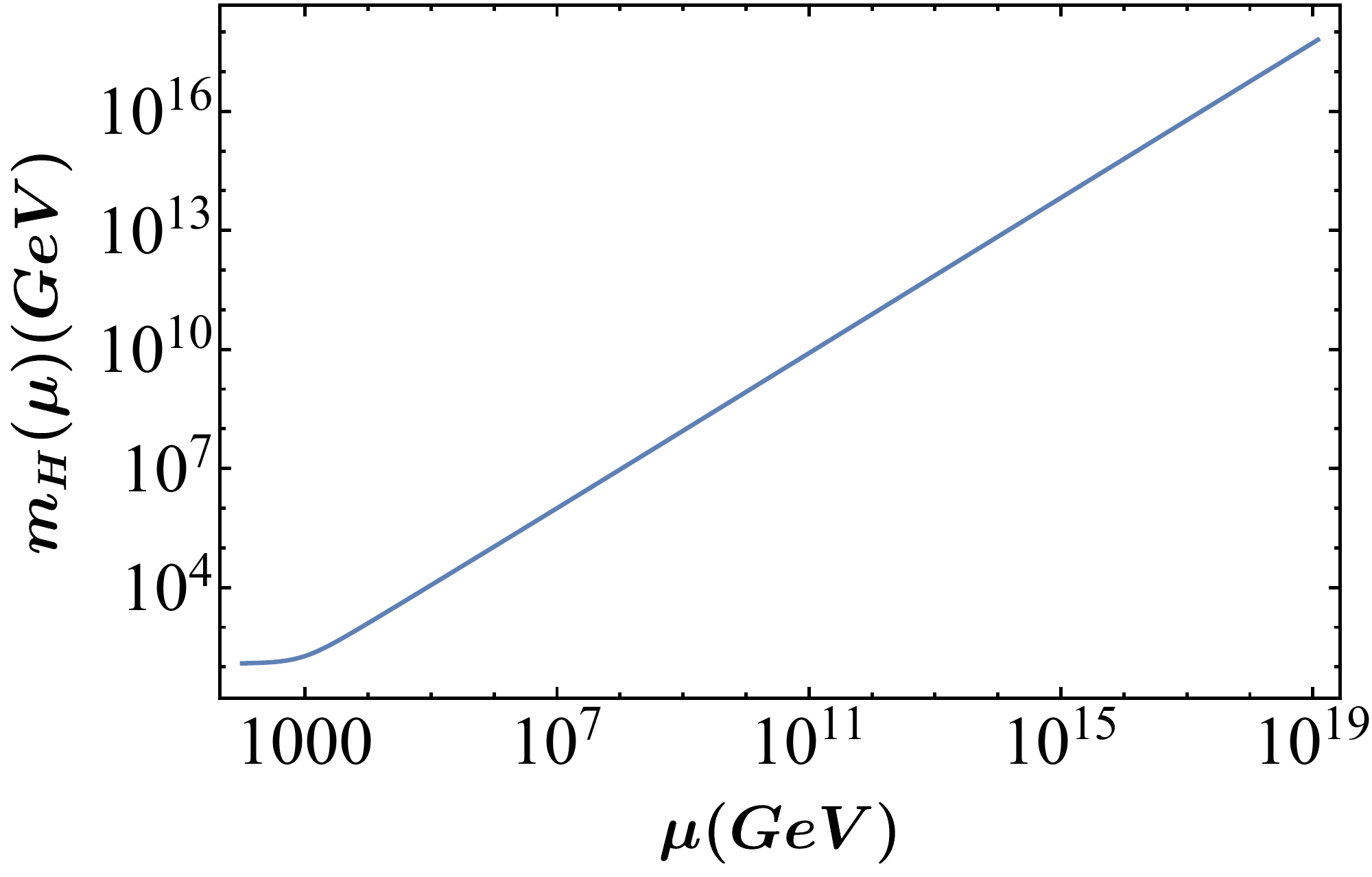}
		\includegraphics[scale=0.41]{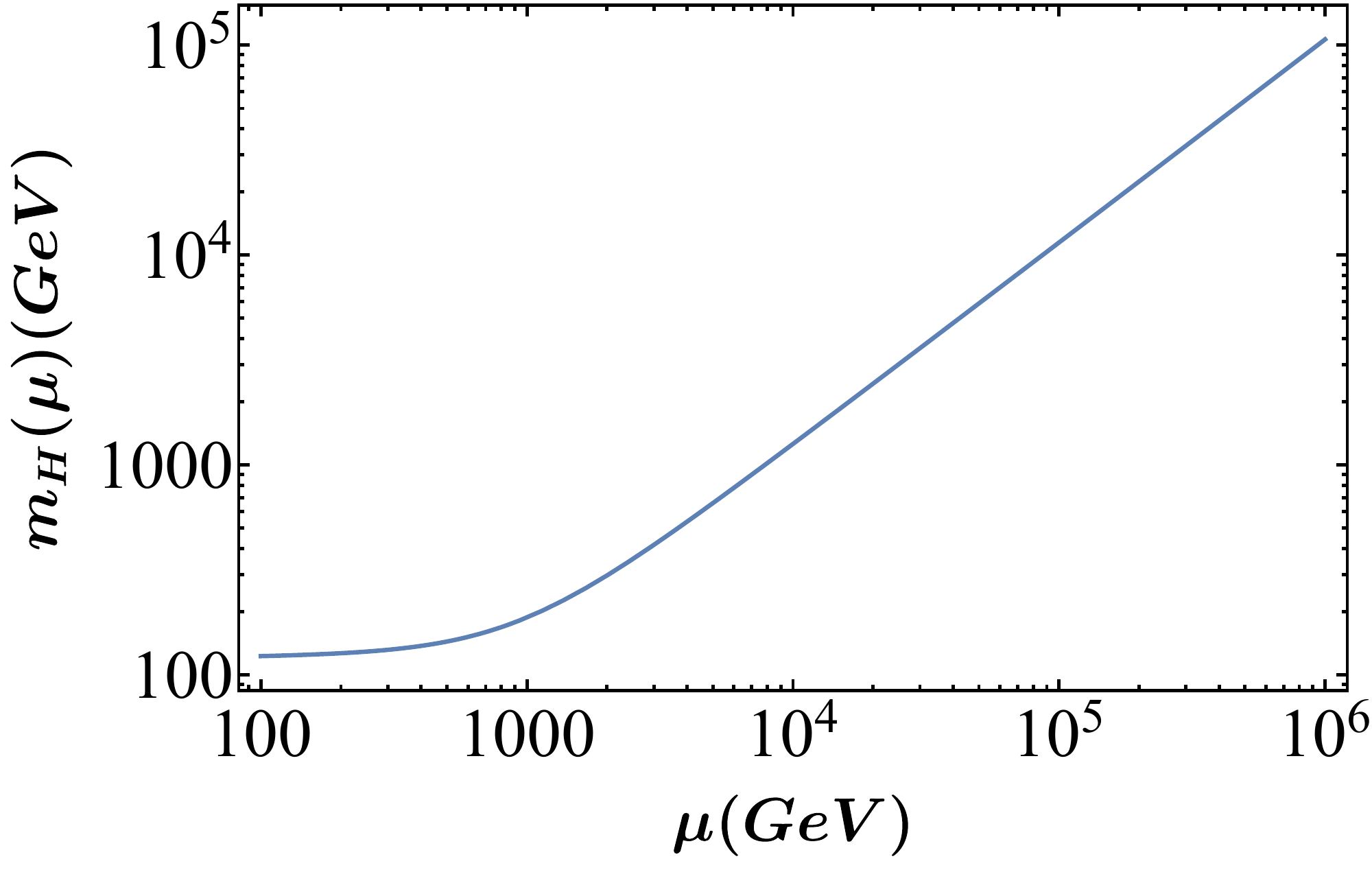}
\caption{{\it Left panel}: Log-log plot of $m_{H}(\mu)$, with UV boundary  $m_H(M_P) \sim 6.347 \cdot 10^{17}$ GeV, see Eq.\,\eqref{wilsmass}. The latter  is coupled to the RG equations for the SM couplings, $\lambda$, $y_t$, $g_1$, $g_2$ and $g_3$, solved numerically  using one-loop beta functions, and IR (experimental) boundary values: 
$\lambda(m_t)=0.1272$, $y_t(m_t)=0.9369$, 
$g_1(m_t)=0.3587$, $g_2(m_t)=0.6483$, $g_3(m_t)=1.1671$ ($m_t$ is the top quark mass). Eq.\,\eqref{approxmSM} (analytical approximation to the solution of\,\,\eqref{wilsmass}) is also plotted, but the two curves are indistinguishable. 
{\it Right panel}: Zoom  in the region $10^2 - 10^6$ GeV of the running shown in the left panel. The \vv elbow'' around $\mu \sim 10^3$ GeV signals that the IR flow is entering the region where $m_{H}(\mu)$  is very well approximated by $m_{H, r}(\mu)$ (Eq.\,\eqref{solh}).}

	\label{Running Standard Model 2}
\end{figure*} 

There is a third important lesson contained in  (\ref{riscrittura}), that is related to the following question. 
Should we identify the physical running mass with (\ref{riscrittura}) or with   (\ref{solpert_RG}), i.e. with the original Wilsonian mass $m^2_\mu$, or with the subtracted mass  $m^2_{\mu, r}$? 
In QFT the running mass is usually identified with (\ref{solpert_RG}). On the other hand, according to the  definition of Wilsonian action, the running couplings $g_i(\mu)$ at the scale $\mu$ result from the integration over the quantum fluctuations in the range $[\mu, \Lambda]$, and are the effective couplings at this scale. This is true, in particular, for the mass. 
Therefore, it is the original Wilsonian mass $m^2_\mu$, not  the subtracted   $m^2_{\mu, r}$ that has to be identified with the physical mass at the scale $\mu$. 

Being this the case, how can we justify the traditional (textbook) approach to QFT, where it is $m^2_{\mu, r}$ that is identified with the physical running mass at the scale $\mu$? 

The answer to this question comes from
the comparison between our result \eqref{riscrittura} and (the textbook) equation \eqref{solpert_RG}. As long as we confine ourselves to  sufficiently low values of $\mu$ (IR regime), 
the flow governed by (\ref{solpert_RG}) practically coincides with the flow  (\ref{riscrittura}). This overlap region is defined by the condition
\begin{equation}\label{condition}
\frac{\lambda\mu^2}{32\pi^2-\lambda} \ll 
	\left(\frac{\mu}{\Lambda}\right)^{\frac{\lambda}{16\pi^2}}\left(m^2_{_{\Lambda}}+\frac{\lambda \Lambda^2}{32\pi^2-\lambda}\right),
\end{equation}
where (we stress again) the term inside the parenthesis contains the fine-tuning necessary to obtain the IR (measured) value of the physical mass. 
Moreover, our equation \eqref{riscrittura} allows to find the energy range to which  Eq.\,\eqref{solpert_RG} is limited. Clearly, if we are interested in energy scales $\mu$ above the region determined by \eqref{condition}, we must go back to the original flow \eqref{riscrittura}, that has a much wider range of validity. 

We are now ready to move to the SM, and to present our proposal (a similar scenario holds if extended versions of the SM are considered). Following steps similar to those that led to (\ref{runningeqm}), for the Higgs mass we get the Wilsonian RG equation 
\begin{align}
	\label{wilsmass}
	\mu\frac{d}{d\mu}\,m^2_H&=\alpha(\mu)\,\mu^2+\gamma(\mu)\,m^2_H,
\end{align}
where $\alpha(\mu)$ is a combination of SM running couplings (gauge, Yukawa, 
scalar), that at one-loop level reads (for our purposes it is sufficient to restrict ourselves to the one-loop order) 
\begin{equation}
16\pi^2 	\alpha(\mu)=12 y_t^2-12\lambda-\frac{3}{2}\,g_1^2-\frac{9}{2}\,g_2^2,
\end{equation}
and $\gamma(\mu)$ is the mass anomalous dimension
\begin{equation}
	16\pi^2 \gamma(\mu)=6 y_t^2+12\lambda-\frac{3}{2}\,g_1^2-\frac{9}{2}\,g_2^2.\,\,\,
\end{equation}

Considering (as for the scalar theory above) constant values for the couplings, from (\ref{wilsmass}) we obtain
\begin{equation}\label{approxmSM}
	m_H^2(\mu)=\left(\frac{\mu}{\Lambda}\right)^{\gamma}\left(m_H^2(\Lambda)-\frac{\alpha\,\Lambda^2}{2
		-\gamma}\right)+\frac{\alpha\,\mu^2}{2
		-\gamma}\,,
\end{equation}
that, as it is easy to check, provides a very good approximation to the flow governed by \eqref{wilsmass}. An improvement to \eqref{approxmSM} is obtained if   $\alpha$ and $\gamma$ 
outside the parenthesis 
are replaced with $\alpha(\mu)$ and $\gamma(\mu)$ (the term in parenthesis is an integration constant, where $\alpha=\alpha(\Lambda)$ and $\gamma=\gamma(\Lambda)$). 
In Fig.\,1 both the solution to \eqref{wilsmass} and its analytical approximation \eqref{approxmSM} (with the improvement mentioned above)
are plotted. They are practically indistinguishable.

Eq.\,\eqref{approxmSM} is one of the most important result of the present work, and deserves several comments. Before doing that, however, it is worth to derive few other related results. Let us define (as in (\ref{scalarMC})) the critical mass,
\begin{equation}
	m_{H,\rm cr}^2(\mu)\equiv \frac{\alpha \,\mu^2}{2-\gamma}\,, 
\end{equation}
and (as in \eqref{mtilde}) the subtracted mass
\begin{equation}\label{mhr}
	m_{H, r}^2(\mu) \equiv m_{H}^2(\mu) -  m_{H,\rm cr}^2(\mu)\,.
\end{equation} 
From \eqref{approxmSM} we derive the equation
\begin{align}\label{renoeq}
	\mu\frac{d}{d\mu} m_{H
		, r}^2(\mu)&=\gamma\,\, m_{H,r}^2(\mu)\,,
\end{align}
that once solved gives ($\mu_0\, > \mu$)
\begin{equation}\label{solh}
	m^2_{H, r}(\mu)=\left(\frac{\mu}{\mu_0}
	\right)^{\gamma}
	m^2_{H, r}(\mu_0)\,.
\end{equation}

Eq.\,(\ref{renoeq}) coincides with the well-known (textbook) one-loop improved RG equation for the renormalized running mass, and is nothing but  Eq.\,\eqref{submass}. We then conclude that $m_{H, r}^2(\mu)$  defined in (\ref{mhr}) {\it is} the usual renormalized running Higgs mass.
However, going back to \eqref{approxmSM}, we observe that (as explained in the section devoted to the scalar theory) it is the Wilsonian mass parameter $m_H^2(\mu)$ that has to be identified with the running Higgs mass. In connection with that, let us consider now the two following things.

(i) If we require that  $m_H^2(\mu)$ at the Fermi scale $\mu_F$ is the measured $m_{H, exp}^2 \sim (125.7)^2$\,GeV$^{2}$, from  \eqref{approxmSM}
we see that $m_H^2(\Lambda)$ needs to be enormously fine-tuned. 

(ii) Turning to the RG flow \eqref{solh} for $m_{H, r}^2(\mu)$, and this time we require that it is   
$m^2_{H, r}(\mu_F)$   that takes the experimental value 
$\sim (125.7)^2$ GeV$^2$, we see that  the two flows $m_H^2(\mu)$ and $m_{H, r}^2(\mu)$ {\it coincide} for all the values of $\mu$ that satisfy the condition  
\begin{equation}\label{cond2}
	\frac{\alpha\,\mu^2}{2
		-\gamma} \ll 	\left(\frac{\mu}{\Lambda}\right)^{\gamma}\left(m_H^2(\Lambda)-\frac{\alpha\,\Lambda^2}{2
		-\gamma}\right) \,.
\end{equation} 

These results contain crucial physical lessons. 
First of all we learn that the \vv fine-tuning''of $m_H^2(\Lambda)$, that in the  traditional approach to QFT is {\it formally} realized through the introduction of counterterms,
has a profound  {\it physical} meaning. It provides the boundary at the UV scale $\Lambda$ for the RG flow of the running mass $m_H^2(\mu)$. A very large value of $m^2_H$ at the UV scale $\Lambda$  is {\it physically necessary} and welcome, not an unwanted result to get rid of. 

Moreover, from Eq.\,\eqref{approxmSM} and from Fig.\,1, we see that through a {\it quadratic running}  that lasts for most of the $m_H^2(\mu)$ flow towards the IR, this finely tuned value of  $m_H^2(\Lambda)$
allows to reach the experimental value of the Higgs mass  at the Fermi scale. What is crucial to realize is that, proceeding towards the IR,  the initial \vv quadratic running''  $m^2_H(\mu) \sim \mu^2$ sooner or later gives the way to a lower energy running, where the \vv multiplicative renormalization''  (see (\ref{solh})) emerges. In schemes as DR we only  have access to (\ref{solh}), but from a truly  physical perspective the latter is an \vv emergent property'' of the running, that rises when the flow approaches the IR. 

This is a great change in the usual paradigm. In the physically unavoidable top-down Wilsonian approach,  
a large hierarchy between the UV and the IR values of $m_H^2$, together with the fine-tuning of $m_H^2(\Lambda)$, are {\it physically mandatory} and the typical multiplicative renormalization (\ref{solh}) emerges as an IR property of the complete  physical running \eqref{approxmSM}. 

Moreover Eq.\,\eqref{approxmSM}  shows the limitations of \eqref{renoeq}, or equivalently of its solution \eqref{solh}: they  can be used only at sufficiently low energies (where  \eqref{cond2} is satisfied), that in most of the cases are the only experimentally reachable energies. 
In the right panel of Fig.\,1, a zoom of the $m_H(\mu)$ running is shown. The presence of an \vv elbow" near $\mu \sim 10^3$ GeV and the \vv freezing" of the  $m_H(\mu)$ flow at lower scales  signal that the first term in the right hand side of \eqref{approxmSM} takes over the second one. This realizes the \vv transition'' from the additive to the multiplicative renormalization of the mass, that is the 
transition from \eqref{approxmSM}  to \eqref{solh}. If one were to extend \eqref{solh} outside its realm of validity, the experimental value $m_H = 125.7$ GeV at the Fermi scale would be reached starting with the UV boundary $m_H(M_P)\sim 132.4$ GeV. This is connected with one of the popular (but, as shown above, incorrect) approaches to the NH problem (see Eq.\,\eqref{natur} and the related discussion).

The fine-tuning manifests itself through the term $(m_H^2(\Lambda)-\alpha/(2-\gamma)\Lambda^2)$   in \eqref{approxmSM}, and the choice of this combination in the UV determines the measured (IR) value of $m^2_H$. Therefore, taking into account the experimental uncertainties, we conclude that there exists a region of \vv tiny size'' in the SM parameter space from which very large UV boundary values of $m^2_H$ give rise, through the RG flow, to the measured (within errors) value of the Higgs mass. 
Such a region can only be inherited from the ultimate UV completion of the SM (or of the yet unknown BSM), namely the Theory of Everything. In String Theory, for instance, an enormous variety of theories/vacua has to be considered, and the conditions for the existence of such a region are certainly met. Moreover, in connection with the quadratic dependence of the Higgs mass in the UV, we note that in the string framework the common expectation is that  the Higgs mass at the string scale $M_S$ is $m^2_H(M_S)\sim M_S^2$ \,\cite{Abel:2021tyt}.    

Going back to the running \eqref{solh} (multiplicative renormalization), we observe that it can be obtained within different schemes (DR, heat kernel, ...), and that no  physical content can ever be related to the choice of a specific scheme. However our analysis has shown that this behavior is related
{\it only} to the IR sector of the flow. The whole UV $\to$ IR running is given by the Wilsonian flow \eqref{approxmSM}, while \eqref{solh} is confined to the IR regime alone. 
\begin{figure}[t]
	\centering
	\includegraphics[scale=0.42]{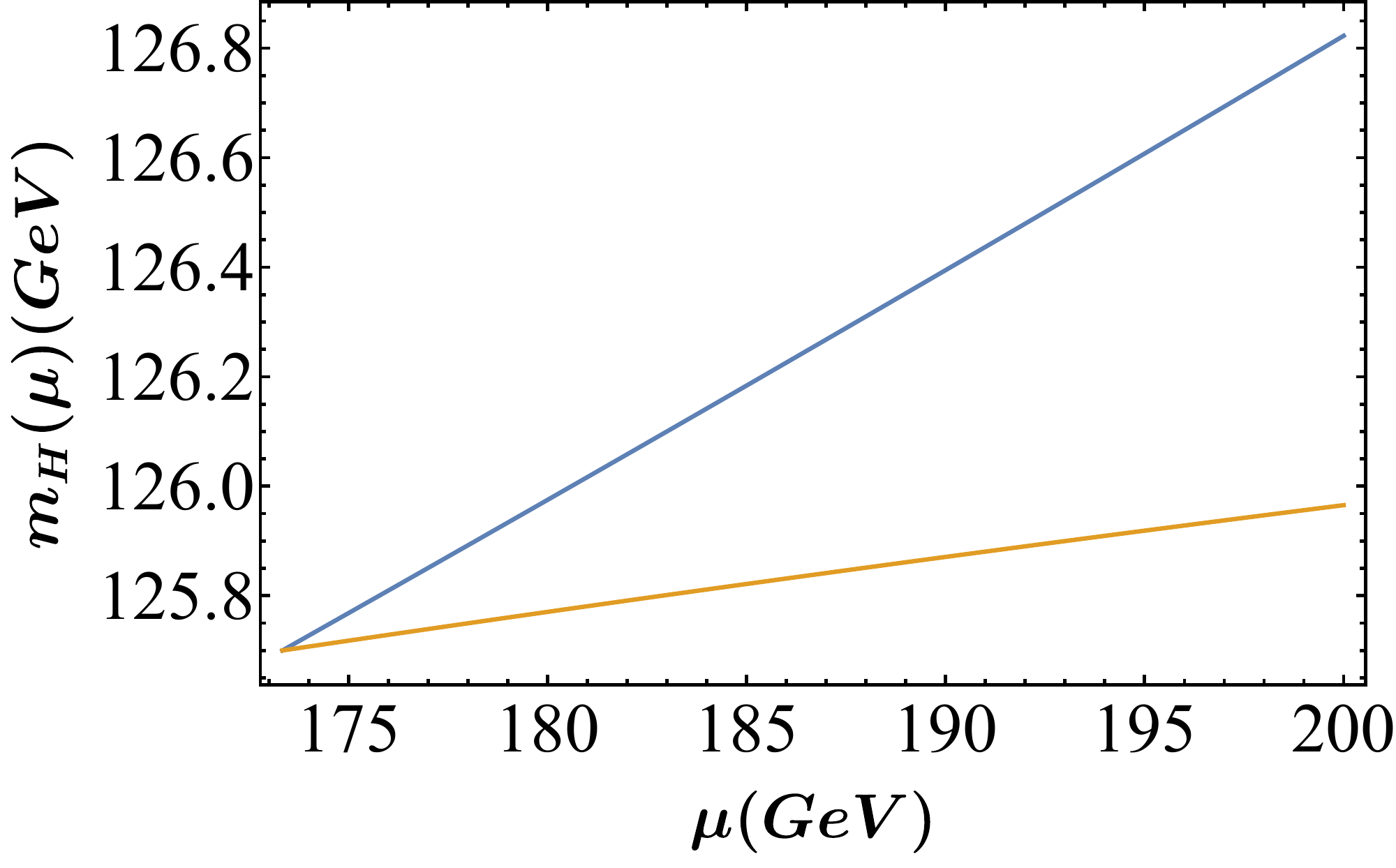}
	\caption{This figure shows a focus of the $m_H(\mu)$ flow of Fig.\,1, Eq.\,\eqref{approxmSM}, in the IR region between $m_t$ (where $m_H(m_t) \sim 125.7$ GeV) and $200$ GeV (blue line). The yellow curve is the flow given by \eqref{solh} again with $m_H(m_t) \sim 125.7$ GeV. Future experiments should allow to evidentiate the difference between these two flows. } 
	\label{Freezing in the IR SM}
\end{figure}

In light of these findings, an interesting question arises, that might be  subject to experimental investigation in the (hopefully not too far) future. Although no one has observed up to now the running of the Higgs mass, we can consider physical processes that should allow to test the $m^2_H(\mu)$ flow (much in the same spirit of what is done with the running bottom quark  mass\,\cite{Aparisi:2021tym}): think, for instance, of ongoing work on precision measurements of the trilinear  coupling\,\cite{Degrassi:2019yix}.  If future experiments will be able to enter the energy regime where the complete flow \eqref{approxmSM} and the approximate IR flow \eqref{solh} start to be  significantly different, and experimentally distinguishable, it should become possible to discriminate between these two alternatives (see Fig.\,2).  

In this respect, we observe that the connection between quantum field theory and statistical physics is usually done  by establishing a one-to-one correspondence between the request $\xi \gg a$ in the theory of critical phenomena ($a$ is the lattice spacing, $\xi$ the correlation length) and 
the request  $m^2 \ll \Lambda^2$ in the QFT framework ($m$ is the particle mass, $\Lambda$ the ultimate UV scale of the theory).When phrased in RG language, this corresponds to the tuning towards the \vv critical surface'', achieved through the subtraction of the \vv critical mass'': $m^2_{ren}(\mu)=m^2(\mu)-m^2_{cr}(\mu)$.
However, we have seen that $m^2_{ren}(\mu)$ only captures the {\it final part} of the running of the physical mass. Actually, the RG flow is physically meaningful even far from the critical surface and from fixed points. This is in fact what happens to our flow \eqref{approxmSM}, that approaches the critical surface of the gaussian fixed point in the IR, giving eventually rise to the flow \eqref{solh}. 

These points can be well illustrated if we go for a moment to $d=3$ dimensions, and consider the Ginzburg-Landau free energy (used to describe the ferromagnetic transition), $F[\phi]=\int d^3 x \left(\frac12(\grad{\phi})^2 + V_k(\phi)\right)$, with potential  
$V_k(\phi)=\frac 12 m^2_k\phi^2 +\frac{\lambda_k}{4!}\phi^4$. 
The RG equations 
for the dimensionless couplings $\widetilde m_k^2\equiv k^{-2} m_k^2$ and $\widetilde \lambda_k\equiv k^{-1}\lambda_k$ ($t \equiv \log k/k_0$, with $k_0$ a reference scale) are
\begin{align}\label{eqmassa}
	\frac{d\widetilde m^2_k}{dt}&=-2 \widetilde{m}^2_k-\frac{\widetilde \lambda_k }{4 \pi^2 (1+\widetilde m^2_k)}\\
	\label{eqcoupling}
		\frac{d\widetilde \lambda_k}{dt}&=- \widetilde{\lambda}_k+\frac{3 \widetilde \lambda_k^2 }{4 \pi^2 (1+\widetilde m^2_k)^2}\,.
\end{align}
It is immediate to see that these equations have a Gaussian and a Wilson-Fisher fixed point, G and WF in Fig.\,3, and that  G  is an IR repulsive fixed point.

\begin{figure}[t]
	\centering
	\includegraphics[scale=0.238]{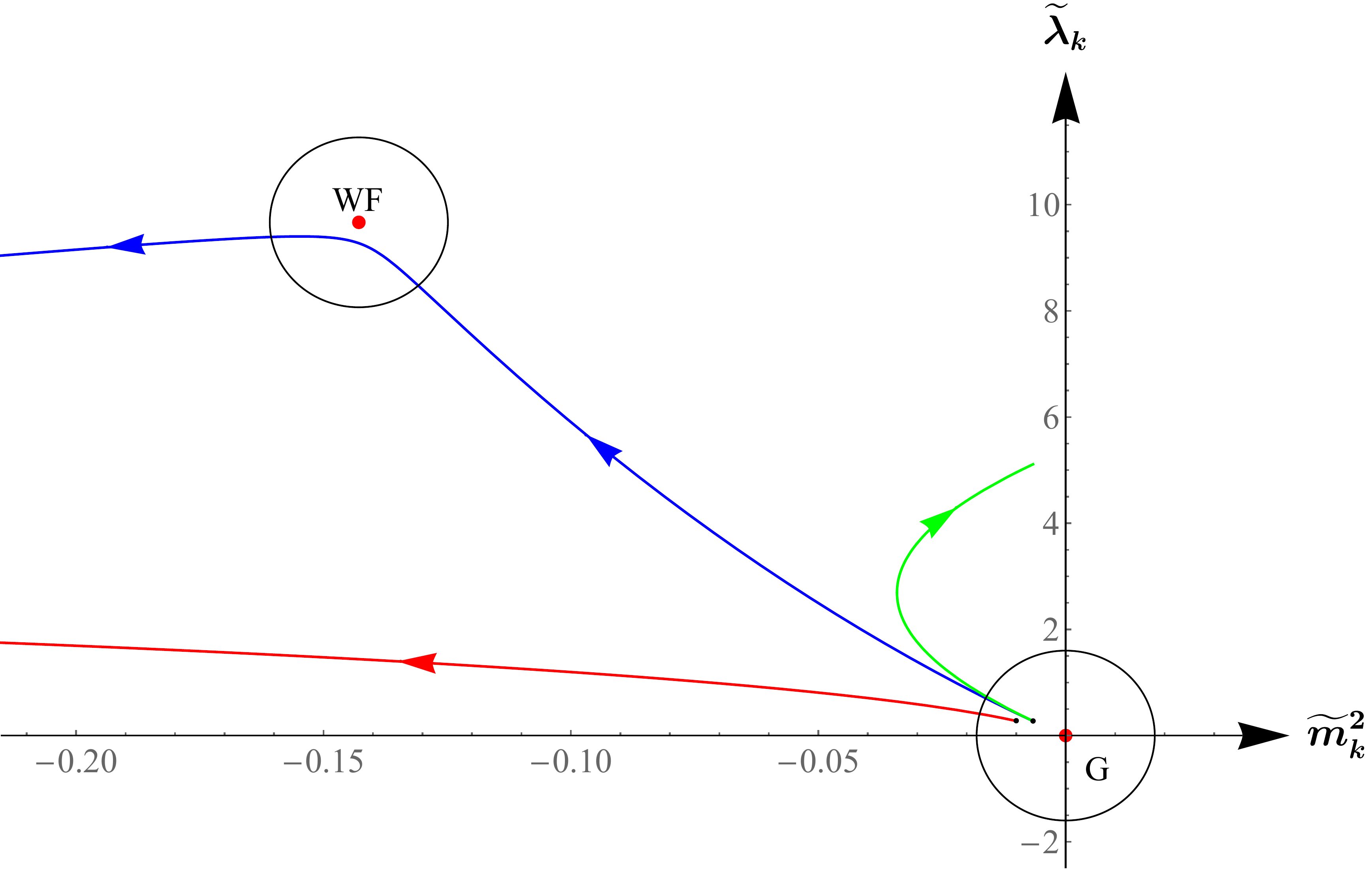}
	\caption{RG flows  (Eqs.\,\eqref{eqmassa} and \eqref{eqcoupling}) in the parameter space $(\widetilde m_k^2,\widetilde\lambda_k)$ of a $\phi^4$ theory in $d=3$ dimensions. The blue and red flows emanate from the UV region close to the Gaussian fixed point G (different boundary values). The green line is obtained linearizing \eqref{eqmassa} and \eqref{eqcoupling} around G, with the same boundary as the blue one.} 
	\label{Freezing in the IR SM}
\end{figure}

Fig.\,3 conveys two messages. ({\bf a}) - Let us consider the UV $\to$ IR flow given by the blue line. In the region around G where \eqref{eqmassa} and \eqref{eqcoupling}  can be linearized, this flow  is well approximated by the \vv subtracted flow'' (green line), the analog of \eqref{solh} in this case. Beyond this region, however, the green flow deviates from the true physical flow (blue line). The very existence of the ferromagnetic transition shows that the green flow {\it cannot} be the true one. ({\bf b}) - The  blue and red flows have slightly different UV boundaries. Thanks to the fine-tuning operated in the UV, the blue flow is driven towards WF (i.e.\,towards the ferromagnetic transition). This example clearly shows that, if (as it is certainly the case) the IR physics is dictated by WF, the fine-tuning in the UV is {\it physical} and {\it unavoidable}. 

\section*{Acknowledgments}

We would like to acknowledge M. Bochicchio, N. Darvishi, M. Grazini, F. Quevedo, D. Zappalà for useful discussions.

\end{document}